# Core-clad Pr(3+ )-doped Ga(In)-Ge-As-Se-(I) glass fibers: Preparation, investigation, simulation of laser characteristics

E.V. Karaksina [1], V.S. Shiryaev [1, *], M.F. Churbanov [1], E.A. Anashkina [2], T.V. Kotereva [1], G.E. Snopatin [1]

[1] G.G. Devyatykh Institute of Chemistry of High-Purity Substances of RAS, 49 Tropinin Str., 603950, Nizhny Novgorod, Russia
[2] Institute of Applied Physics of RAS, 46 Ulyanov Str., 603950, Nizhny Novgorod, Russia
* E-mail address: shiryaev@hps.nnov.ru (V.S. Shiryaev).

a b s t r a c t s

Accepted 5 July 2017

Core-clad fibers on the basis of high-purity 1300 ppmw Pr(3+ )-doped Ga-Ge-As-Se and 2000 ppmw Pr(3+ )-doped In-Ge-As-Se-I glasses have been prepared. The minimum optical losses of the fibers are at the level of 1 dB/m in the wavelength range, where there is no influence of the praseodymium absorption. At present, it is the best result obtained for core-clad rare-earth-doped chalcogenide fibers with high dopant concentration. The fibers exhibit the broadband luminescence in the spectral range of 3.5 - 5.5 μm. The values of luminescence lifetime at the wavelength of 4.7 μm are within 6.5- 8.2 ms closed to those for the bulk glasses obtained previously. Simulation of laser properties for these core-clad fibers in the three-level cascade scheme was carried out. The low level of optical losses experimentally achieved in these fibers is one of main criteria for development of mid-IR fiber systems.

## 1. Introduction

Chalcogenide glasses (ChG) doped with rare earth elements (REE) are widely discussed in the scientific literature regarding the possibility of their using for creation of middle infrared (mid-IR) fiber lasers and amplifiers. At present, laser fiber materials for the spectral region beyond 3.9 μm have not been experimentally developed. So, ChGs are of current interest due to their characteristics, such as low phonon energies (down to 350 cm$^{-1}$ ), high transparency in the mid-IR range, potential ultra low optical losses, an ability to dissolve the rare earth elements, as well as the high calculated values of the quantum efficiency and lasing slope efficiency.

To characterize the laser fiber materials, the quantum efficiency determined as a ratio between experimental and calculated values of emission lifetime, the lasing slope efficiency, and the threshold power are used to be considered. These characteristics have been evaluated by numerical simulations for the number of chalcogenide multi-component glasses doped with REE, taking into consideration the different pump parameters [1- 6]. Data showed the high values of quantum efficiency (up to 100%) and the lasing slope efficiency (up to 54%) of these glasses for the mid-IR range.

The main advantages of selenide chalcogenide glasses, as potential laser media, are low phonon energy values (230-300 cm$^{-1}$ ), the possibility to incorporate the rare earth elements in relatively high concentration (thousands of ppm), high values of refractive index (2.6- 2.9, depending on glass composition), and, as a consequence, the high values of absorption cross section at the pump wavelength.

The set of these properties of ChGs had predetermined a choice of research subjects here, namely, Pr(3+ )-doped Ga(In)-Ge-As-Se glasses and fibers as the materials promising for mid IR-range.

Due to the need to prepare fiber laser materials with low optical losses, the main focus had been done on the development of ChGs with a low content of limiting impurities which have negative influence on the spectral characteristics of glasses in the range of pump and emission wavelengths. The aim of this research was to fabricate the core-clad fibers on the basis of high-purity 1300 ppmw Pr(3+ )-doped $Ga_3Ge_{17}As_{18}Se_{62}$ and 2000 ppmw Pr(3+ )-doped $Ge_{15}As_{16}Se_{63}In_3I_3$ core glasses, which could satisfy to requirements for the laser fiber materials, first of all, high dopant concentration and low optical losses, as well as to perform the simulation of fiber laser characteristics by using the experimental values of the fiber properties.

## 2. Background

Here, we give a comparative analysis of the available literature data on the characteristics of doped selenide ChGs in terms of their potential as active fiber sources of mid-IR radiation. Table 1 presents recent data on the calculated and experimental properties of REE-doped ChGs, which characterize these materials as laser media. High values of emission lifetimes at different mid-IR wavelengths experimentally achieved for a number of glass compositions and the high quantum efficiencies demonstrate potential ability of these materials for development of mid-IR lasers. In addition, REE-doped ChG fibers are attractive for development of emission fiber sources and amplifiers for the mid-IR. For example, the simulations [7,8] have shown, that for ChG fibers with compositions closed to ones considered here, the high pump-to-signal slope efficiencies can be achieved. In particular, for the Pr(3+)-doped Ga-Ge-As-Se glass fiber, it can reach 45% at a wavelength of 4.5 μm [8].

Simulation of laser characteristics for REE-doped ChG fibers, for different cavity configurations, has demonstrated the high potential for development of these materials for mid-IR fiber lasers, including tunable ones [3,5,6,9]. For the three-level cascade scheme, lasing properties of REE-doped core-clad ChG fibers were estimated. Calculated values of the pump-to-signal slope efficiency for these Pr(3+), Dy(3+) and (Tb(3+)–doped core glass fibers were within 8–54%, depending on the cavity type, the pump wavelength and fiber characteristics [3,5,6].

The possibility of experimental realization of the potential of REE-doped ChGs as materials for mid-IR fiber lasers and amplifiers is determined by their real characteristics that could be achieved in accordance with the design parameters. The last experimental data showed the progress in this field, nevertheless laser systems on the basis of ChG fiber materials have not been developed yet.

Pr(3+)-doped glasses are of special interest due to the ($^3F_2$, $^3H_6$) → $^3H_5$ and $^3H_5$ → $^3H_4$ radiative transitions, associated with relaxation of the excited states of praseodymium atoms, and to form the wide-broad emission spectrum in the range of 3.5–5.5 μm, leading to their potential use for creation of tunable fiber lasers and broadband light sources. As can be seen from Table 1, Pr(3+)-doped ChGs possess the highest values of quantum efficiency and emission lifetime.

It is known that, for doped ChGs and optical fibers, radiated in the spectral range of 3–6 μm, the main problem is the presence of limiting impurities in the forms of the Se-H, S-H and Ge-H groups (absorption bands are at 4.57, 4.01 and 4.9 μm, respectively), as well as carbon- and oxygen-contained impurities in the forms of - OH groups, C-C-bonds, $H_2O$, $CO_2$, $CS_2$, $CSe_2$, arsenic and germanium oxides. These impurities provide the high optical losses in fibers as negative factor. The locations of their absorption bands are close to the pump and emission wavelengths in the spectra of the doped ChGs. The process of removal of these limiting impurities is very complicated. For optical fibers, with increasing of optical path-length of the radiation propagation (at wavelengths of pump and emission), the requirements for the permissible content of these impurities are more critical. Obviously, reduction of their concentration would be good factor for increasing of the radiation efficiency. In this regard, the preparation of low loss core-clad ChG fibers with required core/clad parameters, optical and emission characteristics and high pump-to-signal efficiencies are actual tasks.

According to the simulation of laser performances for these REE-doped fiber materials, the one of criteria of their use for fiber lasers is low optical losses (less than 1 dB/m) at the pump and emission wavelengths [1–3,5,7,8]. Reducing of optical losses can be achieved by using the purification methods such as chemical distillation techniques [10,11], allowed to obtain high purity glasses in relation to dissolved impurities and inclusions formed as a result of glass crystallization and/or as incomplete dissolution of dopants.

To solve the problem of preparing REE-doped ChG fibers with low optical losses, we had developed a multi-stage method for preparation of Ga(In)-Ge-As(Sb)-Se-(I) glasses and fibers with low content of limiting impurities [12–15]. These results were used here for preparation of core-clad doped fibers, their characterization, and evaluation of their potential laser properties based on the known simulation model.

## 3. Glass and fiber preparation

To prepare the core-clad fibers, the $Ga_3Ge_{17}As_{18}Se_{62}$ and $Ge_{15}As_{16}Se_{63}In_3I_3$ glasses had been taken as the host matrix. The synthesis processes of the same glasses in high-purity state were described in Refs. [12–15]. For these core glasses, the clad glass compositions were chosen taking into account their viscosity values, to felicitate the core-clad fiber drawing using a double crucible technique, and the refractive indices, to obtain the required numerical aperture of the fiber. Therefore, for this task, the high-purity $Ga_{2.5}Ge_{17.5}As_{15}Se_{65}$ and $Ge_2As_{39}Se_{59}$ clad glasses had been prepared. Two different clad glasses were used to prepare the core-clad fibers with different numerical apertures.

The synthesis of high-purity Pr(3+)-doped Ga-Ge-As-Se and In-Ge-As-Se-I core glasses is the multi-stage process. The first stage was the same for both glass compositions. It includes the preparation of the Ge-As-Se glass purified by chemical distillation process [10,16,17]. The steps of purification of the gallium and indium compounds by using the chemical vapor transport reversible reactions were described in Refs. [12,14]. The next stages were: the vacuum loading of the pre-purified Ga (or InI) batch to the Ge-As-Se glass one; co-melting and homogenization of these mixtures; loading of $Pr_2S_3$ (5N purity) batch to the host glass one under dry Ar atmosphere; the vacuum melting and homogenization of these mixtures in the rocking muffle furnace at 750–800 °C for 10–12 h; subsequent quenching in air, annealing at $T_g$ and slow cooling. The values of Pr(3+) concentration in the different core glasses were 1300 and 2000 ppmw.

**Table 1**
Values of emission lifetime and quantum efficiencies for doped ChGs and fibers.

| Materials | Dopant elements | Dopant concentration, ppmw | Emission lifetime, ms (4.0–4.8 μm) | Quantum efficiency, % | Reference |
|---|---|---|---|---|---|
| GaGeAsSe (bulk) | Pr | 350–1500 | 10–12 | 80–100 | [1,2,4,5,8] |
| | Dy | 1000–2000 | 6–6.1 | 97 | |
| | Tb | 500–1500 | 11–11.8 | 73 | |
| GaGeAsSe (fiber) | Pr | 500 | 7.8 | – | [19,21] |
| GaGeSbS (bulk) | Er | 10000 | 0.72 | 64 | [4] |
| GaGeSbS (fiber) | Er | 1000 | (emission spectrum) | – | [4] |
| InIGeAsSe (fiber) | Pr | 1300 | 4 | – | [13] |
| InIGeSbSe (bulk) | Pr | 1300 | 6 | – | [22] |

The $Ga_{2.5}Ge_{17.5}As_{15}Se_{65}$ clad glass was obtained by melting of high-pure $Ga_3Ge_{17}As_{18}Se_{62}$ and $GeSe_4$ glasses pre-synthesized using of chemical distillation purification technique. The $Ge_2As_{39}S_{59}$ clad glass was prepared by melting of mixture of a high-pure $As_4S_4$, elemental sulfur and germanium (in the required ratio of elements) in evacuated silica-glass ampoule in a rocking muffle furnace. To reduce the content of hydrogen, oxygen, and carbon particles as limiting impurities, the purification methods described in Refs. [10,16] were used.

The core and clad glass samples were prepared in the form of rods of diameter 8 and 12 mm and length 40–50 mm.

Mono-index and core-clad fibers were drawn using a single and a double crucible technique [10]. There were no visible traces of glass crystallization during the fiber drawing. The core-clad fibers were coated with a protective fluoroplastic F-42 layer of 10–20 μm thickness. Tens of meters of glass fibers (with different core and clad glass compositions and various core/clad diameter ratio) were prepared, and their optical losses were measured using the conventional cut-back technique.

## 4. Properties of glasses and fibers

The quantitative elemental compositions of the prepared glass samples were determined by energy dispersion X-ray microanalysis using the scanning microscope SEM-515 (Philips, Netherlands) equipped with an energy-dispersive detector EDAX 9900 (EDAX, USA). The error of composition measurements was (0.3–0.5) at.%.

Thermal analysis (DSC) of the glass samples was carried out using a synchronous Netzsch STA 409 PC Luxx analyzer with sensitivity of 1 μV/mW and temperature accuracy ± 1K, at the heating rate of 10 K/min in the temperature range 50–550 °C. The DSC curves for the core and clad glass samples were characterized by the only endothermic peak corresponded to $T_g$, that demonstrates the absence of liquation during the preparation process. The $T_g$ values for 1300 ppmw Pr(3+)-doped $Ga_3Ge_{17}As_{18}Se_{62}$ and 2000 ppmw Pr(3+)-doped $Ge_{15}As_{16}Se_{63}In_3I_3$ core glasses, were 228 °C and 195 °C, respectively. The $T_g$ values for $Ga_{2.5}Ge_{17.5}As_{15}Se_{65}$ and $Ge_2As_{39}S_{59}$ clad glasses were 175 °C and 206 °C, respectively. It is important to note, that the DSC curves of all doped and undoped glass samples did not exhibit crystallization exothermic signals up to 550 °C that shows their low tendency to crystallize. It is the new result for ChGs concerning the possibility to achieve a relatively high dopant concentration without of crystallization of glass fiber during the drawing process.

Glass and fiber samples were analyzed using IR-spectroscopy (Fourier transform IR spectrometer Tensor 27) in the spectral range 350−7000 cm⁻¹.

Fig. 1 shows the absorption spectra of undoped and doped glass bulk samples: 1 - $Ga_3Ge_{17}As_{18}Se_{62}$ (host glass); 2 - $Ge_{15}As_{16}Se_{63}In_3I_3$ (host glass); 3 - $Ga_{2.5}Ge_{17.5}As_{15}Se_{65}$ (clad glass); 4 - 1300 ppmw Pr(3+)-doped $Ge_3Ge_{17}As_{18}Se_{62}$ (core glass); 5 - 2000 ppmw Pr(3+)-doped $Ge_{15}As_{16}Se_{63}In_3I_3$ (core glass). The spectrum of $Ge_2As_{39}S_{59}$ clad glass is not shown due to an absence of impurity absorption bands in the spectral range 7000-1000 cm⁻¹.

The optical losses of these fibers were measured by the conventional cut-back technique. The measured fiber lengths were within 0.1–1 m. To reduce cladding modes, the Ga-In alloy immersion of input and output fiber surfaces was applied. Spectra of total optical losses of some prepared mono-index and core-clad glass fibers are given in Figs. 2 and 3.

The total optical losses of the mono-index $Ga_3Ge_{17}As_{18}Se_{62}$ and $Ge_{15}As_{16}Se_{63}In_3I_3$ host glass fibers (Figs. 2 and 3 curves 1) were lower than 1 dB/m in the spectral ranges of 2.5–3.3 μm and 5.5–7.5 μm. There were no impurity absorption bands, except for Se-H bands at 4.57 μm. The minimum optical losses of the core-clad ChG fibers were at the same level in the spectral ranges, where there were no influence of praseodymium absorption. But, in the contrast with undoped fibers, oxygen-contained impurities were present in these fiber samples.

Based on optical fiber loss values (Figs. 2 and 3), the concentration of hydrogen and oxygen impurities was determined using the extinction coefficients ε: ε(Se-H) = 1780 cm⁻¹/wt.% at the wavelength of 4.57 μm, ε(Se-H) = 100 cm⁻¹/wt.% at 3.5 μm [10]; ε(As-O) = 2 cm⁻¹/wt.% at 7.46 μm, ε(As-O) = 6 cm⁻¹/wt.% at 7.9 μm [18]. The values of limiting impurity concentration for fiber samples are given in Table 2. For clad glasses (two lower rows), these concentrations were determined for bulk samples (in accordance with data of Fig. 1).

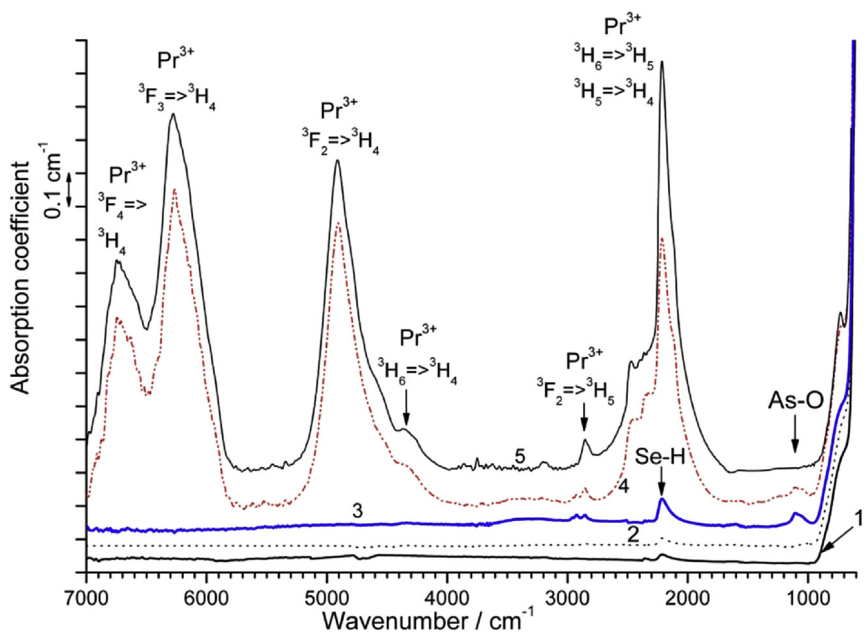

**Fig. 1.** Absorption spectra of glasses: 1 - $Ga_3Ge_{17}As_{18}Se_{62}$ and 2 − $Ge_{15}As_{16}Se_{63}In_3I_3$ (host glasses); 3 - $Ga_{2.5}Ge_{17.5}As_{15}Se_{65}$(clad glass); 4 - 1300 ppmw Pr(3+)-doped $Ga_3Ge_{17}As_{18}Se_{62}$ and 5 - 2000 ppmw Pr(3+)-doped $Ge_{15}As_{16}Se_{63}In_3I_3$ (core glasses).

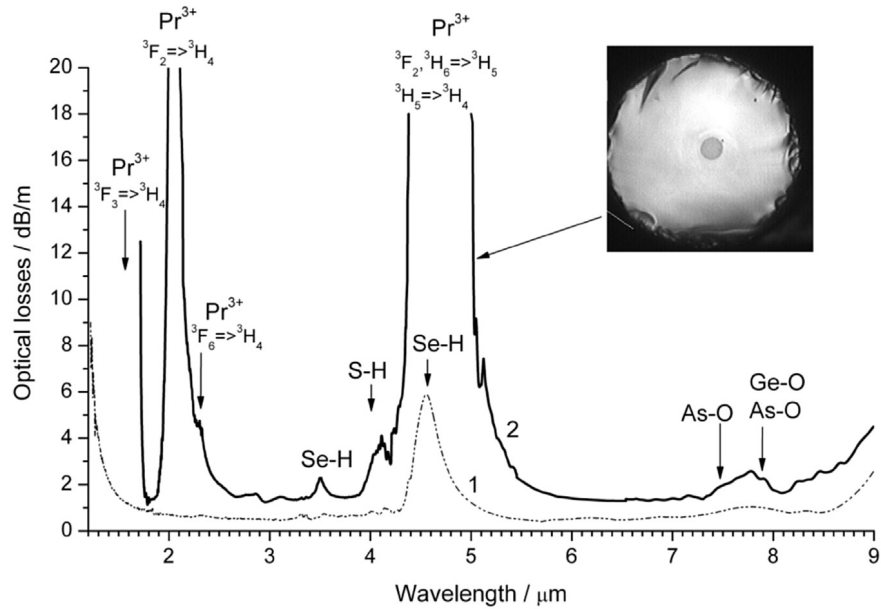

**Fig. 2.** Optical losses of mono-index Ga$_3$Ge$_{17}$As$_{18}$Se$_{62}$ glass fiber (1); core-clad 1300 ppmw Pr(3+)-doped Ga$_3$Ge$_{17}$As$_{18}$Se$_{62}$/Ga$_{2.5}$Ge$_{17.5}$As$_{15}$Se$_{65}$ glass fiber (2). Insert is cleave photo of the fiber 2 (d$_{core}$ = 36 μm, d$_{clad}$ = 330 μm).

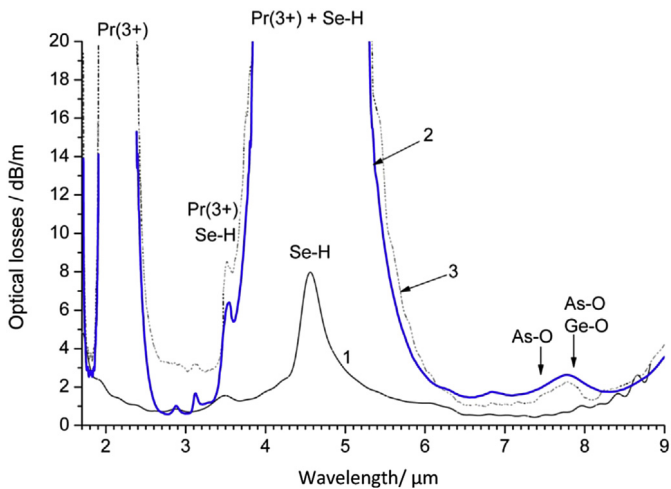

**Fig. 3.** Optical losses of mono-index Ge$_{15}$As$_{16}$Se$_{63}$In$_3$I$_3$ host glass fiber (1), mono-index 2000 ppmw Pr(3+)-doped Ge$_{15}$As$_{16}$Se$_{63}$In$_3$I$_3$ glass fiber (2); core-clad 2000 ppmw Pr(3+)- Ge$_{15}$As$_{16}$Se$_{63}$In$_3$I$_3$/Ge$_2$As$_{39}$Se$_{59}$ glass fiber (3), d$_{core}$ = 80 μm, d$_{clad}$ = 220 μm.

**Table 2**
Concentrations of limiting impurities in the fiber samples.

| Glass composition | Impurity content, ppmw | |
|---|---|---|
| | hydrogen | oxygen |
| Ga$_3$Ge$_{17}$As$_{18}$Se$_{62}$ | 0.07 | <0.5 |
| 1300 ppmw Pr- Ga$_3$Ge$_{17}$As$_{18}$Se$_{62}$ | – | 6 |
| Ge$_{15}$As$_{16}$Se$_{63}$In$_3$I$_3$ | 0.08 | <0.5 |
| 2000 ppmw Pr- Ge$_{15}$As$_{16}$Se$_{63}$In$_3$I$_3$ | – | 3 |
| Ga$_{2.5}$Ge$_{17.5}$As$_{15}$Se$_{65}$ (clad glass) | 0.3 | 10 |
| Ge$_2$As$_{39}$S$_{59}$ (clad glass) | 0.5 | – |

Doped bulk and fiber sample spectra (Figs. 1–3) contain the Pr(3+) ion absorption bands associated with radiative transitions:

$^3F_4 \rightarrow {}^3H_4$ (1.48 μm), $^3F_4 \rightarrow {}^3H_5$ (2.16 μm), $^3F_3 \rightarrow {}^3H_4$ (1.59 μm), $^3F_3 \rightarrow {}^3H_5$ (2.4 μm), $^3F_2 \rightarrow {}^3H_4$ (2,03 μm), $^3F_2 \rightarrow {}^3H_5$ (3.57 μm), $^3H_6 \rightarrow {}^3H_4$ (2.34 μm), $^3H_6 \rightarrow {}^3H_5$ (4.64 μm) and $^3H_5 \rightarrow {}^3H_4$ (4.73 μm) [1,19,20]. Due to overlap of the praseodymium bands at wavelengths of 3.57, 4.64 and 4.73 μm and the absorption bands of Se-H bond (at 3.5 and 4.57 μm), the hydrogen content in the doped fibers had not been evaluated.

Luminescence spectra of the doped fibers were registered at room temperature by means of IR Fourier spectrometer with InSb detector, cooled by liquid nitrogen. The single-mode Tm-fiber laser (λ$_p$ = 1.97 μm) was used as the pump source (rated pump is 2 W). Emission spectra were measured using long pass filter with the cut-on wavelength of 3 μm. The emission lifetime at 4.7 μm was evaluated on the basis of an exponential decay of signal intensity in time. The error of measurement was 0.5 ms. Luminescence spectra of core-clad fibers are shown in Fig. 4. The photoluminescence observed in the spectral range of 3.5–5.5 μm is associated with radiative Pr(3+) transitions of ($^3F_2$, $^3H_6$) → $^3H_5$ and $^3H_5$ → $^3H_4$ given a main contribution. Absorption bands in this region are caused by CO$_2$ molecules, due to spectra recording in air, and by Se-H impurity absorption typical for all samples. Emission lifetimes for fiber samples at 4.7 μm were within 6.5–8.2 ms. There are no significant differences in emission lifetimes for the fibers of the different core/clad glass compositions.

## 5. Simulation of laser characteristics of the prepared fibers

Simulation of laser characteristics for the three-level cascade scheme (Fig. 5a) was carried out using the model developed for similar materials before [3,5,9]. It is based on numerical solving the self-consistent system containing equations for steady-state ion populations, equations describing excited state absorption and stimulated emission rates, and equations for the evolution of the pump, signal and idler powers. We chose a scheme for the fiber cavity [5] with two pairs of fiber Bragg gratings (FBG) tuned to the signal and idler wavelengths in compliance with Fig. 5b and with the data for reflection coefficients taken from Ref. [3]. The experimental data for optical losses and emission lifetime of the prepared

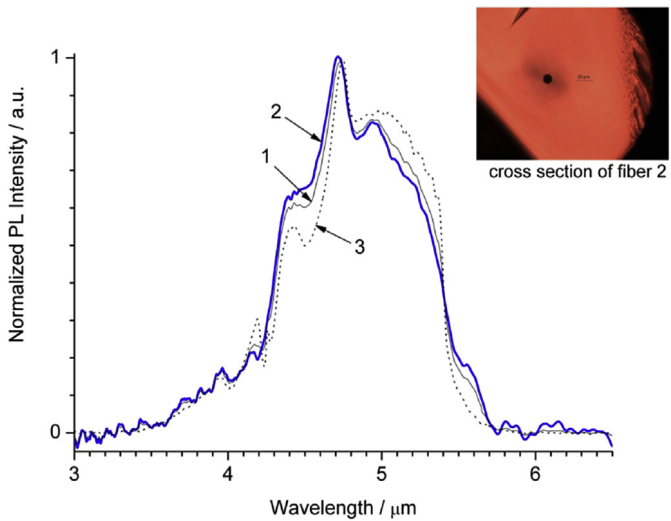

cross section of fiber 2

**Fig. 4.** Photoluminescence spectra of core-clad fibers: 1300 ppmw Pr(3+)- doped Ga$_3$Ge$_{17}$As$_{18}$Se$_{62}$/Ga$_{2.5}$Ge$_{17.5}$As$_{15}$Se$_{65}$ (1); 1300 ppmw Pr(3+)-doped Ga$_3$Ge$_{17}$As$_{18}$Se$_{62}$/Ge$_2$As$_{33}$S$_{59}$ (2); 2000 ppmw Pr(3+)-doped Ge$_{15}$As$_{18}$Se$_{63}$In$_3$I$_1$/Ge$_2$As$_{39}$S$_{59}$ (3), normalized at the wavelength of 4.7 μm. Insert is the cleave photo of the fiber 2 (d$_{core}$ = 16 μm, d$_{clad}$ = 320 μm).

fibers were used in modeling. The reflectivity of the input FBG for the pump wavelength was 0.05 while for the signal and the idler ones it was 0.95. The output reflectivity was 0.05 for the signal and 0.9 for the idler wavelengths. This scheme (Fig. 5a) corresponds to effective depopulation of the N$_3$ upper level with a long lifetime due to the laser operation at the idler wavelength. The cascade scheme allows to reduce the parasitic thermo-optic effect due to radiative transition between N$_3$ and N$_2$ levels.

To calculate the fiber laser characteristics, we used experimental data for core-clad fibers, namely, dopant concentration, optical losses, numerical aperture of fibers, the core/clad ratio, and emission lifetime values. The pump wavelength λ$_p$ was 1.97 μm. The idler wavelength λ$_i$ (Fig. 5a) was 3.8 μm; the signal wavelengths λ$_s$ were close to the emission band maximum. We have considered three values of λ$_s$ as signal ones: 4.5, 4.75 and 5 μm, which correspond to different emission σ$_{se}$ and absorption σ$_{sa}$ cross sections for Pr(3+) ions, shown in Fig. 6 a.

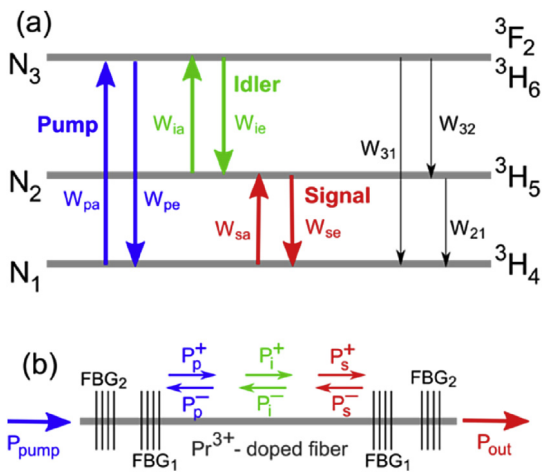

**Fig. 5.** a- Scheme of Pr(3+) energy levels with indicating the radiative (W$_{pa}$, W$_{pe}$, W$_{ia}$, W$_{ie}$, W$_{sa}$, W$_{se}$) and non-radiative (W$_{21}$, W$_{32}$, W$_{31}$) transitions used in simulation procedure; b - Scheme of cascade fiber laser. Bragg gratings FBG$_1$ and FBG$_2$ are adjusted to the signal and idler wavelengths.

The pump-to-signal slope efficiency (η) and threshold power (P$_{th}$) at the different wavelengths in the range of emission, assuming fiber losses of 1 dB/m, which are close to experimental ones, are given in Fig. 6a. Numerical results were obtained to be very good approximated by P$_{out}$ = (P$_{pump}$ - P$_{th}$)·η/100%. The dependences of the output lasing power P$_{out}$ on the optical losses at the different emission wavelengths are shown in Fig. 6b. We calculated the output power at each signal wavelength for different pump powers P$_{pum}$ and fiber lengths L.

Calculated values of output lasing power were carried out for a single-mode beam propagated along the fiber core [21]. It should be noted that a single-mode operation for the core-clad fibers prepared here is possible in principle, by using the mode selection methods (cooling of the fiber, special resonator configurations, tapering of the fiber and some others). However, a detailed analysis of optimum laser design for realization of single-mode operating is beyond this paper.

## 6. Results and discussion

In this paper, we report the preparation of core-clad fibers on the basis of high-purity 1300 ppmw Pr(3+)-doped Ga-Ge-As-Se and 2000 ppmw Pr(3+)-doped In-Ge-As-Se-I glasses having the high dopant concentration, the minimum optical losses at the level of 1 dB/m, the intensive broadband luminescence in the spectral range of 3.5−5.5 μm, and encouraging high calculated laser characteristics.

According to Table 2, the content of the limiting impurities in the doped glass fibers is higher as compared with host ones, which indicates the contaminating effect of the Pr$_2$S$_3$ dopant. It should be noted that the content of gas-forming impurities in the Pr$_2$S$_3$ had not been controlled. The contaminating effect occurs in all fibers independent of the glass composition. The optical losses (Figs. 2 and 3) due to oxygen-containing groups (As-O and Ge-O) in the spectra of the doped glass fibers and the concentrations of impurities (Table 2) are higher by several times as compared with undoped samples. Apparently, an increase of the hydrogen content in the doped glass fibers in the form of Se-H groups took place as well, although their concentrations had not been determined due to the overlap of Se-H and Pr(3+) absorption bands. In addition, in spectra of the doped glass fibers, the absorption bands of S-H bonds had appeared, that is the result of incorporation of the sulfur-component of the Pr$_2$S$_3$ dopant from uncontrolled hydrogen content in the glass network. Therefore, the procedure of purification of the Pr$_2$S$_3$ dopant from limiting gas-forming impurities is urgent. However, the Pr$_2$S$_3$ compound is characterized by high melting temperature and low vapor pressure over the melt, which hinders its distillation purification and requires a special purification technique.

The minimum optical losses of the doped glass fibers in the spectral ranges of 2.7−3.4 and 6−7.5 μm (where the influence of absorption praseodymium is minimum) are at the level of 1 dB/m (Figs. 2 and 3). This result satisfies to requirements which have been developed for fiber laser sources. As can be seen from Figs. 2 and 3, the transmission spectra of mono-index host glass fibers (curves 1) demonstrate the lower level of optical losses in a wide spectral range, that confirms the absence of the contamination of fiber glasses at the stage of fiber drawing. First of all, this result is important for glass purification from the hydrogen impurity as hard-removing one. Its concentration in fiber samples is at the level of 0.1 ppmw, that is close to values for host bulk glasses [13,14]. These data are also important in terms of reproducibility of the method for preparation of high-pure doped glass fibers with low optical losses in the range of pumping and emission. At the same time, unfortunately, there are problems of dopant purification and additional optical losses due to an imperfection of the core-clad fiber boundary. The analysis of these problems is beyond this paper and requires a special study.

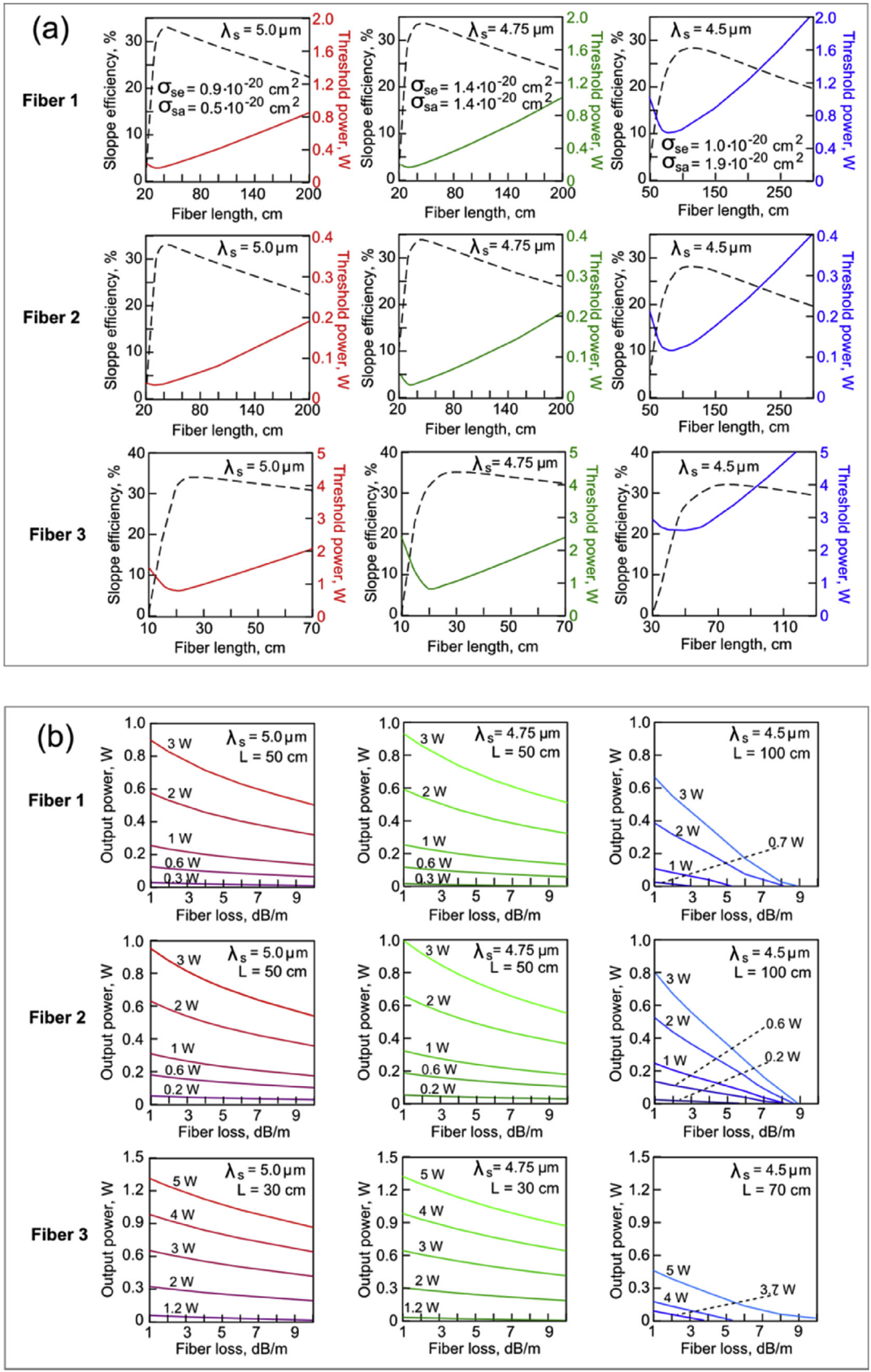

**Fig. 6.** a- Dependence the slope efficiency (dashed line, left axis) and the pump power threshold (solid line, right axis) on the fiber length with different glass compositions (optical loss is 1 dB/m). Fiber 1 - 1300 ppmw Pr(3+)-doped $Ga_3Ge_{17}As_{18}Se_{62}$, $d_{core} = 36$ μm, clad- $Ga_{2.5}Ge_{17.5}As_{15}Se_{65}$; Fiber 2 - 1300 ppmw Pr(3+)-doped $Ga_3Ge_{17}As_{18}Se_{62}$, $d_{core} = 16$ μm, clad - $Ge_2As_{30}S_{59}$; Fiber 3 - 2000 ppmw Pr(3+)-doped $Ge_{15}As_{16}Se_{63}In_3I_3$, $d_{core} = 80$ μm, clad -$Ge_2As_{39}S_{59}$, b- Dependence of the output power $P_{out}$ at the signal wavelength on the values of optical loss at different pump powers ($\lambda_p = 1.97$ μm).

It is important to note, that the experimental results obtained here demonstrate the lowest optical losses and highest Pr(3+)-ions concentration in core-clad REE-doped chalcogenide glass fibers in comparison with the world's data for this type of glass materials [1,23]. This result is the new step on the way of development of mid-IR fiber lasers and amplifiers.

Prepared mono-index and core-clad fibers, pumped by Tm-fiber laser (1.97 μm), exhibit the broadband luminescence in the spectral range of 3.5—5.5 μm (Fig. 4). Luminescence spectra for the fibers are close to those recorded previously for the bulk glass samples with the same composition. The values of emission lifetime at the wavelength of 4.7 μm for core-clad fibers are within 6.5—8.2 ms, which is a little lower than the previously published data for the bulk samples with the same composition (8—11 ms). Likely, it is due the loss of the radiation extended beyond the fiber core [22]. It can be noted, that the emission band shape is not dependent of the glass composition that is in correlation with the data published previously. The more high value of praseodymium concentration (up to 2000 ppmw) in comparison with the typical published data for the doped ChG fibers was achieved due to high dopant solubility in the InGeAsSeI matrix [14]. At 2000 ppmw Pr(3+) concentration, the crystallization of glasses during the fiber drawing process was not observed, that had been confirmed by the spectral characteristics of the fibers: the transparency of several meters long fibers, and closeness of optical losses of doped and host glass fibers in the wide spectral range (Fig. 3, curves 1,2). These data demonstrate an advantage of prepared Pr(3+)-doped fibers as compared to ones published before. For example, according to [22—24], the Pr(3+) content in selenide ChGs should be in the range of 500—1000 ppmw to avoid of glass crystallization and/or concentration quenching of luminescence, while we did not observe these effects up to 2000 ppmw concentration of Pr-ions.

Simulation of laser characteristics for prepared fibers was carried out on the basis of the three-level cascade scheme (Fig. 5). Calculations have demonstrated the promising laser characteristics of these fibers (Fig. 6). The values of pump-to-signal slope efficiency are high, up to 33—35% at wavelengths of 4.5, 4.75 and 5 μm. The pump power thresholds $P_{th}$ were different for fibers. The fiber 2 had a lower $P_{th}$ value in comparison with fibers 1 and 3. This value is lower than known optical damage threshold for ChGs, therefore fiber 2 can be potentially more attractive laser material. Maximum value of output power (Fig. 6a) would be achieved for the short lengths of fibers (50 and 30 cm) for different glass compositions. For these ones, the optical loss effect would be negligible. Nevertheless, dependencies of output power on the optical losses at the wavelengths in the range of emission band (Fig. 6b) show that, with increasing of optical losses by ten times, the output power reduces by 40% independent on the glass composition. The results are of interest in terms of estimation of the influence of undesirable absorption due to hard-removing impurities, such as hydrogen, on the potential fiber laser characteristics.

## 7. Conclusions

The Ga-Ge-As-Se and In-Ge-As-Se-I glasses, doped with Pr(3+) ions had been prepared and studied. On the basis of these glasses, the fibers with different core and clad glass compositions had been prepared by the double-crucible method. The minimum optical losses in the core-clad doped glass fibers were about 1 dB/m in the wavelength ranges of 2.7—3.4 μm and 5.7—7.5 μm. All samples of the optical fibers exhibited an intensive broadband luminescence in the spectral range of 3.5—5.5 μm. The values of emission lifetime at the wavelength of 4.7 μm for fiber samples were within 6.5—8.2 ms. Low level of optical losses and high Pr(3+) dopant concentration had demonstrated an advantage of prepared fiber samples in comparison with published data for chalcogenide fiber materials promising for the development of mid-IR lasers and amplifiers. By using the experimental properties of these fibers, the simulation of potential laser characteristics for wavelengths, corresponding to bandwidth of emission, had been performed.

## Acknowledgments

The authors express their sincere gratitude to prof. I.A. Bufetov for useful comments and discussion of the results. The experimental work is supported by the Russian Science Foundation (Grant No. 16-13-10251). The numerical simulation provided by E.A. Anashkina is supported by the Russian Foundation for Basic Research (Grant No 16-32-60053).